\documentclass[sigconf,screen]{acmart}

\ifdefined\directlua
  \pdfvariable gentounicode=1
  \pdfextension glyphtounicode {a117}{2709}
\fi

\makeatletter
\def\UrlBigBreaks{\do@url@hyp}

\makeatother

\usepackage{listings}
\usepackage{multirow}
\usepackage{xspace}

\copyrightyear{2026}
\acmYear{2026}
\setcopyright{cc}
\setcctype{by}
\acmConference[ASE '26]{Proceedings of the 41st IEEE/ACM International Conference on Automated Software Engineering}{October 12--16, 2026}{Munich, Germany}
\acmBooktitle{Proceedings of the 41st IEEE/ACM International Conference on Automated Software Engineering (ASE '26), October 12--16, 2026, Munich, Germany}
\acmDOI{10.1145/3832783.3834553}
\acmISBN{979-8-4007-2882-2/2026/10}

\received{2026-05-13}
\received[accepted]{2026-07-02}

\begin{document}

\title{The Ground Is Shifting: A Reflection on the Foundations of Software Measurement}

\author{Thomas Bock}
\correspondingauthor
\orcid{0000-0001-6906-3489}
\affiliation{%
  \institution{Carnegie Mellon University}
  \city{Pittsburgh}
  \state{PA}
  \country{USA}
}
\email{bockthom@cmu.edu}

\author{Audris Mockus}
\orcid{0000-0002-7987-7598}
\affiliation{%
  \institution{University of Tennessee}
  \city{Knoxville}
  \state{TN}
  \country{USA}
}
\email{audris@utk.edu}

\author{Bogdan Vasilescu}
\orcid{0000-0003-4418-5783}
\affiliation{%
  \institution{Carnegie Mellon University}
  \city{Pittsburgh}
  \state{PA}
  \country{USA}
}
\email{vasilescu@cmu.edu}

\renewcommand{\shortauthors}{Bock et al.}

\begin{abstract}
For most of the past six decades, software measurement relied on labor-intensive manual collection of proprietary data, which hampered progress. The shift to repurposing traces from version control and related tools dramatically expanded data availability---especially with the rise of open-source software---but hinged on an often unstated assumption: that these tools are used by professional developers to build genuine software systems.
However, as trace-generating tools, data types and scale, and empirical methods have all evolved, it has become clear that changes in data generation and analytical approaches affect many prior findings about software development, maintenance, and evolution. With AI agents now actively using these same tools, the resulting traces frequently violate the original assumption of human origin. To preserve the relevance of software measurement research, immediate action is needed: We must detect when foundational assumptions are violated in contemporary data and develop new methodologies that remain valid under changed circumstances. To this end, we propose a systematic AI-assisted replication program that revisits key findings using modern techniques, aiming
for methods that yield consistent results on current data to keep software measurement meaningful.
\end{abstract}


\begin{CCSXML}
<ccs2012>
<concept>
<concept_id>10011007.10011074.10011111.10011695</concept_id>
<concept_desc>Software and its engineering~Software version control</concept_desc>
<concept_significance>500</concept_significance>
</concept>
</ccs2012>
\end{CCSXML}

\ccsdesc[500]{Software and its engineering~Software version control}

\keywords{Mining software repositories, squash-merging, AI-generated code}

\maketitle

\section{Introduction}
Empirical software engineering has transformed significantly over the past six decades. The initial notion that software should be measured to inform development became widespread in the 1970s, notably with the introduction of source-code complexity metrics~\citep{mccabe1976complexity}, regression models for developer productivity~\citep{walston1977method}, and early software reliability models~\citep{musa1979software}. These foundational approaches, though impactful, relied on costly manual data collection. The 1980s brought more sophisticated models~\citep{Boehm-81,albrecht1983software},
along with systematic programs for software measurement in enterprises~\citep{grady1987software} and critical evaluations of emerging methodologies~\citep{weyuker1988evaluating}.
Manual data collection remained prevalent and access to data was mostly proprietary.

In the 1990s, empirical software measurement matured further. Researchers proposed frameworks for effective measurement systems~\citep{basili1992software,schneidewind1992methodology},
practical guidance for their application~\citep{putnam1991measures}, and methodologies for qualitative studies~\citep{seaman1999qualitative}. However, measurement continued to be restricted to organizations capable of supporting extensive programs, limiting broader research participation.

A pivotal shift occurred with the realization that valid metrics
could be extracted from projects that use version control and issue trackers. This democratized empirical software engineering: Any project using these systems could analyze its own data and benchmark against historical states, enabling predictions about defect introduction~\citep{mockus2000predicting} and future file reliability~\citep{graves2000predicting}.
Crucially, validation extended beyond proprietary projects~\citep{mockus2002two,mockus2000case},
opening vast public open-source datasets to the research community.
Methodological advances replaced rules for manual metric collection~\citep{kitchenham1995towards} with automated extraction techniques from operational systems.
As platforms evolved---from SourceForge~\citep{howison2004perils} to git's dominance~\citep{bird2009promises} and GitHub's ascendance~\citep{kalliamvakou2014promises}---data extraction approaches were continually refined. Researchers adapted their tools, methods, and datasets in response to rapid technological change and shifting development practices~\citep{pfleeger2025Evidence-Based,fernández2018Empirical,sjøberg2007The}.
The rise of open-source hosting platforms---especially GitHub---dramatically expanded research possibilities. Data could be enhanced by incorporating issue trackers, pull requests (PRs), and continuous-integration logs~\citep{cosentino2017A,cosentino2016findings}.
This enabled analyses at ecosystem scale rather than isolated projects~\citep{boldi2020Ultra-Large-Scale,mockus2007large},
driving the development of mining infrastructure such as GHTorrent or Software Heritage archives~\citep{dabic2024SEART,dueñas2021GrimoireLab,Gousios2012GHTorrent,pietri2020software}.

Recently, datasets have grown to span entire ecosystems. Repository archives such as World of Code aggregate version control data across diverse hosting platforms, constructing unified graphs linking developers, commits, and projects across observable open-source activity~\citep{Ma2019world,pietri2020software}.
These cross-platform resources allow researchers to address previously intractable questions, such as
studying vulnerability propagation via dependencies or examining the diffusion of practices across programming communities~\citep{agroskin2023constructing,ruohonen2025tracing,liu2022demystifying,blanthorn2019evolution,mujahid2023go,he2021large}.
The scale now encompasses billions of artifacts, requiring advanced computational infrastructure and analytics~\citep{boldi2020Ultra-Large-Scale,Ma2019world,pietri2020software}.

Methodological rigor has kept pace: Extracting valid measurements from operational systems is now standard practice,
with greater attention paid to threats to validity, replication standards, and statistical robustness~\citep{ampatzoglou2019identifying,lago2024threats,verdecchia2023threats,sjøberg2021Construct,feldt2010validity}.
Machine learning and \mbox{natural}-language processing techniques are widely adopted---for example, in code completion or vulnerability detection~\citep{lin2020software,durrani2024decade}. \mbox{Qualitative} approaches (e.g., interviews, ethnographies) complement or replace quantitative analyses in many settings~\citep{dybaa2011qualitative,di2017combining,seaman2025qualitative}.

The emergence of large language models (LLMs) and AI-powered development tools marks a new inflection point. Tools such as ChatGPT, GitHub Copilot, or Claude Code are reshaping how developers write, review, and debug code~\citep{osorio2025evaluation,france2024navigating,weber2024significant,santos2025decoding,chatlatanagulchai2025use}.
This evolution challenges established empirical research paradigms: datasets built from human-written code may not reflect current practices; studies validated on pre-AI corpora face questions about ongoing relevance. But also new opportunities arise around understanding developer--AI interaction, measuring the quality impact of generated code, and distinguishing human versus machine contributions. The definition of a ``software artifact'' worthy of study is itself being renegotiated.

Fundamentally, most of the present empirical work is predicated on repurposing operational data (e.g., issue tracking or version control) to measure phenomena such as developer effort or software quality. As processes and tools evolve---particularly with generative AI (GenAI)---the durability of existing measurement methods is uncertain. Methodologies must account for both professional human activity (e.g., keeping logical changes separate in version control) and deviations (e.g., agent-generated commits). Much effort was devoted to cleaning operational data, such as excluding agent activity~\citep{dey2020detecting}
or untangling complex commit histories~\citep{herzig2013impact}. In the age of GenAI agents, extracting valid data
requires major methodological innovation. Recent project practices emphasizing simplified history (e.g., squashed commits) further complicate accurate authorship and evolution tracking. At the same time, some
problems with cleaning data may no longer be relevant, for example, when LLMs generate accurate, standardized commit messages
\citep{zhang2024automatic,xue2024automated,wu2025empirical}.

Urgent action is needed to ensure the survival of
software measurement
as a discipline that repurposes operational data in software projects to gain useful
software-engineering insights. It requires intervention from
tool makers to generate traces that identify automated and manual activities, and from researchers using these
traces to find
ways to
treat
human actions (constrained by physics and biology) and agent actions (not having such constraints).

\section{Assumptions of Data Collection Practices and the Evolution of Analytical Methods}

The evolution of empirical software-engineering research can be understood through two interrelated dimensions: data sources and analytical methods, each of which has undergone substantial change.

Throughout most of its history, software-engineering measurement was hampered by expensive (and limited) manual data collection and proprietary datasets. Only during the last quarter century, the discovery that data from software tools, such as version control systems, could be used to produce valid measures led to exponential expansion in measurement methods, and, in the case of open-source software, in public datasets of millions of projects~\citep{felderer2020The}.
That leap to rely on tool-trace data (i.e., records left by tool usage) required many assumptions---some forgotten by now (see Table~\ref{tab:assumptions}).
Even under these assumptions, trace data had to be augmented for different types of software changes,
outliers identified and removed,
missing data imputed,
and many other contingencies addressed~\citep{mockus2000identifying}.
\begin{table}[t]
  \caption{Foundational assumptions
           and their violations} \vspace*{-0.3cm}
  \label{tab:assumptions}
  \setlength{\tabcolsep}{2pt}
  \scriptsize
  \begin{tabular}{p{4cm}p{\dimexpr\columnwidth-4cm-4\tabcolsep\relax}}
  \toprule
   Assumption & Violation \\
  \midrule
  A commit is an atomic unit of work & Squash-merging collapses PRs into one commit \\
  Traces are generated by human developers & AI agents \& bots generate commits \\
  Version history is append-only & Rebasing \& force-pushing rewrite history \\
  Author identity is unique and stable & Identities are fragmented \& shared accounts exist \\
  \bottomrule
  \end{tabular}
  \vspace{-0.1in}
\end{table}
Numerous violations of
foundational assumptions are already documented~\citep{just2016switching}. For \mbox{example}, \citet{Rapaport2025Altered} showed that altered version histories---through force pushes or rebasing---have become commonplace.
\citet{almarzouq2020mining} emphasized issues of data freshness, API limitations, and the difficulty of establishing ground truth.

While initial regression and software reliability analyses date from the 1970s, most
data analysis methods in software engineering were of a simpler kind, such as descriptive statistics and
correlations.
Gradually, techniques suitable for software data~\citep{de2019evolution}, such as negative binomial and logistic regression, survival analysis, mixed-effects models for hierarchical structure, and Bayesian methods for uncertainty quantification increased in popularity~\citep{felderer2020The,furia2022applying}.
Recently, the field has also begun grappling with causal inference---propensity score matching, instrumental variables, difference-in-differences, and regression discontinuity---recognizing that evaluating interventions without randomized trials requires explicit causal reasoning through directed acyclic graphs~\citep{furia2023towards,graf2024cleaning}.
Network analysis operates at multiple scales: social networks from co-commits or pull requests reveal collaboration patterns~\citep{oliveira2023developers,bock2023automatic}, dependency networks from package managers expose vulnerability propagation~\citep{zerouali2022impact}, and co-change networks illuminate architectural coupling~\citep{zhou2019understanding}. Machine learning has progressed from traditional defect prediction to deep learning on code representations and LLMs~\citep{kabir2025software,abreu2025moving,khalid2023software,wang2023codet5+},
whose integration into workflows (e.g., GitHub Copilot) creates new challenges as
code increasingly reflects AI assistance~\citep{pandey2024transforming}.

Sophisticated methods assume stable relationships, but a model validated on repositories in 2010 may not generalize to 2020 data because the phenomenon itself has changed. \citet{Zimmermann2025Retrospective} identified recurrent challenges: establishing ground truth, overgeneralizing from convenience samples, and the tension between statistical and practical significance. Changes in tools alter both how software is developed and recorded, potentially invalidating assumptions underlying data collection and analytical methods~\citep{Hoess2025does}.
\section{Discussion: Which Findings Are Replicable?}
The preceding sections documented how data sources and analytical methods have evolved over decades of empirical software engineering research,
undermining fundamental assumptions that event traces are generated by a professional human actor following a well-established process with strict quality gates. Would a historic analysis repeated without adjusting for the changes in how the data are now generated lead to the same conclusions?
Of the four assumption violations listed in Table~\ref{tab:assumptions}, we explore two in depth through illustrative anecdotes concerning changes in software development practices,
before turning to the broader implications for the field \mbox{and the potential role of replication and AI in addressing them.}

\subsection{Squash-Merging Hides Developer Activity}\label{sec:squash-merge}

\begin{figure}[t]
\vspace{-0.015in}
\centering
\includegraphics{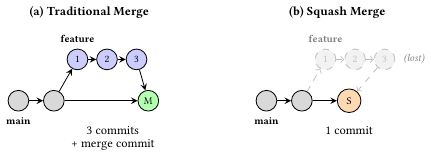}
\vspace{-0.1in}
\Description{Two side-by-side commit graphs comparing traditional merging and squash-merging. On the left, panel (a) shows a traditional merge: a feature branch with three individual commits, numbered 1 to 3, branching off the main branch and rejoining it through a separate merge commit, so all four commits (three feature commits plus the merge commit) remain visible on the main branch. On the right, panel (b) shows a squash merge: the same three feature commits branch off the main branch, but instead of a merge commit, they are collapsed into a single squash-merged commit on the main branch; the three original commits are shown grayed out and dashed, labeled as lost, since they no longer appear in the project history.}
\caption{The same pull request results in different histories: (a)~traditional merging preserves all commits; (b)~squash-merging collapses the commit history into a single commit.}
\label{fig:squash-merge}
\vspace{-0.1in}
\end{figure}

Changes in development practices affect the data empirical researchers analyze. Squash-merging---the practice of collapsing all commits from a pull request into a single commit on the main branch~\citep{Bludau2022pr,kalliamvakou2016depth}---exemplifies how modern workflows can fundamen-\linebreak tally alter what repository mining can observe.
In Figure~\ref{fig:squash-merge}, we illustrate the difference between traditional merge commits and squash merges. In a traditional merge, the feature branch's entire commit history with every single commit is preserved in the main branch, maintaining a granular record of development activity.
In contrast, squash-merging condenses all commits of the pull request into a single commit, erasing intermediate steps.
This practice, while beneficial for maintaining a clean main branch history, has significant implications for empirical research.
Consider what a researcher analyzing commit histories would see when examining a feature implementation under different merge strategies. With traditional merge commits, the history preserves the development trajectory: initial implementation attempts, debugging iterations, responses to code reviews,
and final refinements appear as distinct commits with their own timestamps, authors, and messages. This granular history enables analyses of development processes, bug introduction patterns, and collaboration dynamics.
With squash-merging, this trajectory collapses into a single commit. A feature that required three weeks of development, twenty intermediate commits, and input from multiple reviewers appears as one atomic change attributed to a single author on a single date.
Debugging sessions disappear. Reviewer-prompted modifications vanish.
What remains is a sanitized commit \mbox{that presents the final outcome as if it emerged fully formed.}

The implications for research extend across multiple domains. Bug introduction analysis, which traces defects to the commits that introduced them, cannot pinpoint introductions within squashed commits---the bug and its fix may both be subsumed in the same atomic change~\citep{Rani2024On,Bludau2022pr}.
Developer productivity metrics based on commit counts become meaningless when a month of work is squash-merged into a single commit, obscuring actual effort and activity.
Similarly, analyses of code churn (i.e., frequency and volume of code changes) are distorted when multiple incremental changes are consolidated into one.
Collaboration pattern analysis conducted solely on the project's commit history does not see the changes that might have been conducted during pull-request review.
Also, studies examining the temporal dynamics of development lose fidelity when intermediate timestamps are lost.
Further, developer networks constructed from co-change relationships between commits become sparser and less informative, neglecting the
incremental changes on which the
developers have worked.
If multiple developers collaborated on a feature branch but their contributions are squashed into a single commit, the resulting network \mbox{fails to capture the true collaboration structure.}

\addtolength{\tabcolsep}{-1pt}
\begin{table}[t]
    \caption{Number of commits in merged pull requests (PRs)}\vspace*{-0.3cm}
  \label{tab:merge-comparison}
    \scriptsize
    \begin{tabular*}{\columnwidth}{l@{\extracolsep{\fill}}r@{\hspace{1.167pt}}r@{\hspace{1.167pt}}r@{\hspace{1.167pt}}r}
    \toprule
      Project w.\ squash-merges & \# PRs & Avg. commits/PR & Max. commits/PR & PRs > 1 commits \\
    \midrule
      \textsf{denoland/deno} & 15\,142 & 4.3 & 389 & 61.8\% \\
      \textsf{electron/electron} & 25\,669 & 2.8 & 1\,256 & 38.9\% \\
      \textsf{facebook/react} & 12\,836 & 2.6 & 1\,900 & 35.3\% \\
      \textsf{flutter/flutter} & 48\,051 & 3.6 & 1\,398 & 48.7\% \\
      \textsf{\mbox{huggingface/transformers}} & 19\,353 & 7.3 & 846 & 64.7\% \\
      \textsf{microsoft/vscode} & 47\,094 & 2.8 & 3\,896 & 38.4\% \\
      \textsf{microsoft/TypeScript} & 14\,877 & 3.8 & 438 & 51.7\% \\
      \textsf{twbs/bootstrap} & 9\,259 & 3.3 & 9\,187 & 42.7\% \\
      \bottomrule
  \end{tabular*}
  \par\smallskip\noindent
  \scriptsize Time window: Project start on GitHub until May 12, 2026. Only merged PRs are considered.
\end{table}
\begin{figure}
    \centering
    \vspace*{0.05in}
    \includegraphics[width=0.9\columnwidth]{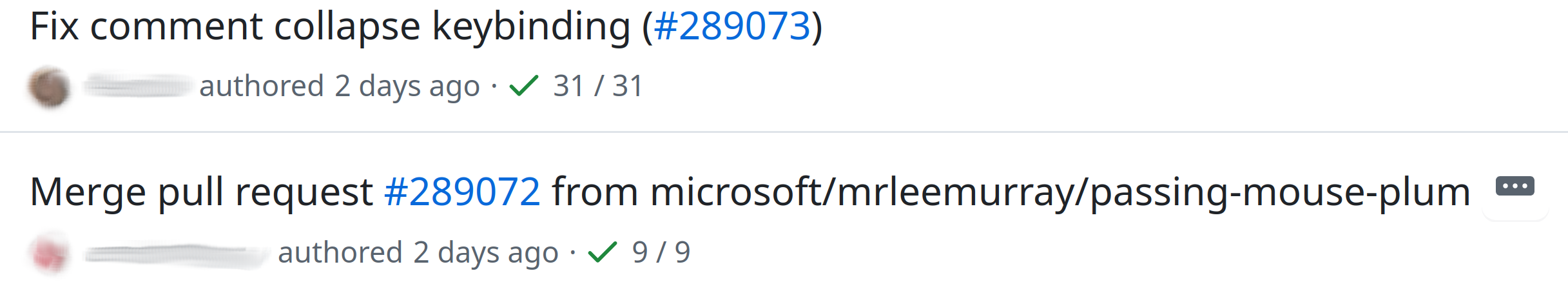}
    \vspace*{-0.05in}
    \Description{Screenshot of a git commit history excerpt from the microsoft/vscode repository, showing two adjacent commits. The lower commit is a traditional merge commit, with a message starting with ``Merge pull request ...''. The upper commit is a squash-merged commit, with a message ending in a pull request number in parentheses, illustrating that both merge styles appear side by side in the same project history.}
    \caption{Excerpt of the commit history of \textsf{microsoft/vscode}, showing a traditional merge commit (lower part) directly next to a squash-merged commit (upper part).}
    \label{fig:merge-vs-squash}
\end{figure}

\begin{table*}[t]
  \caption{Examples of proposed replications targeting human-behavior constructs}\vspace*{-0.3cm}
  \label{tab:replications}
  \setlength{\tabcolsep}{4pt}
  \scriptsize
  \begin{tabular}{p{3.3cm}p{3.4cm}p{\dimexpr\textwidth-3.3cm-3.4cm-6cm-8\tabcolsep\relax}p{6cm}}
  \toprule
   Construct & Classic finding & What to replicate & Why it might break \\
  \midrule
      Productivity \& effort estimation & Commit count \& frequency serve as \mbox{effort proxy}~\citep{robles2022development,streit2024benchmarking,oliveira2020code,shah2023mining} & Analyze commit activity and developer time for pre- and post-GenAI periods & Squash-merged commits as well as AI-generated commits do not reflect development time and human effort \\

      Bug introduction \& prediction & Code churn predicts defects \citep{tahir2023early,li2020systematic,shin2010evaluating} & Measure churn in projects with/without squash-merging and with/without AI assistance & Squash-merging hides change frequency; code churn of AI-generated code may reflect insufficient prompting rather than defects \\
      Developer expertise \& code ownership & Code contributions reflect expertise~\citep{dey2021representation} & Investigate contributions and their quality & Expertise may shift from code writing to reviewing/prompting \\
  \bottomrule
  \end{tabular}
  \vspace{-0.102in}
\end{table*}

The problem compounds because squash-merging adoption is neither universal nor random. Some projects mandate it; others \linebreak forbid it; many leave the choice to individual contributors.
For \mbox{example}, the commit history of project \textsf{microsoft/vscode} currently shows a mixture of traditional merge commits of pull requests (indicated by \emph{``Merge pull request ...''}) and squash-merged pull requests (indicated by \emph{``(\# PR number)''} at the end of the commit message) next to each other, as we show in Figure~\ref{fig:merge-vs-squash}.
Analyses that treat commits as comparable units across such heterogeneous data risk systematic biases whose direction and magnitude depend on the correlation between merge practices and the phenomena under study.
To illustrate the magnitude of the potential problem, we collected descriptive statistics via the GitHub API in May 2026 for 8~widely-used software projects that (at least, partially) use squash-merging. As we show in Table~\ref{tab:merge-comparison}, merged pull requests on average contain 2--7 commits, with maxima up to 9\,187 commits. 35\%--65\% of the pull requests in these projects contain more than one commit and vanish into a single commit when squash-merged.
To assess how prevalent squash-merging is at population scale, we then identified all GitHub projects that have at least one commit whose title follows GitHub's standardized pattern for squash-merged pull requests (i.e., the title ends with \emph{``(\# PR number)''}). Using data from the repository archive World of Code~\citep{Ma2019world} until October 2025, considering only public projects with more than 1\,000 commits (as merging practices primarily matter in projects with large activity), we found that 132\,251 out of 341\,397 projects (39\%) contain at least one such squash-merged commit, indicating that squash-merging is not uncommon.
Project-level adoption of squash-merging rose from 18\% in 2016 to 39\% in 2025, indicating a growing trend.
Our data collection even undercounts squash merges, as project owners can customize the format of commit messages for squash merges, making automatic identification of all squash merges even more complicated.
To the best of our knowledge, past studies have not sufficiently investigated \mbox{how their results are dependent on these merge practices.}

\subsection{Human-Generated vs.\ AI-Generated Data}

Squash-merging already demonstrated that development practices can erase authentic history.
Whereas this erasure is a side effect of a workflow choice,
an AI agent can create a sequence of commits with realistic timestamps, messages, and diffs, which also does not reflect genuine human development.
Even human commits are now hybrid artifacts---partly human intent, partly AI generation.
When a developer uses GitHub Copilot to write a function and then commits it, the trace looks identical to a purely human-authored commit~\citep{xiao2026self}, but the underlying generative process is fundamentally different. The distinction between ``human'' and ``artificial'' is no longer binary---it is a spectrum that current tooling cannot resolve.
With a wider deployment of GenAI agents that use not just version control, but also
conduct code reviews, etc.,
historically held assumptions about humans generating all important software-engineering events no longer hold~\citep{treude2025,robbes2026agentic}. At least some software produced entirely via prompts would no longer be of primary interest for human-behavior constructs, but the corresponding prompts might be~\citep{Watanabe2026}.
Consequently, without a significant change, the current way of \mbox{software measurement via trace data would no longer be relevant.}

\subsection{A Call for Action}

These anecdotes illustrate a broader concern: classic
findings may not be repeatable (without adjustments) on contemporary data that do not satisfy the same assumptions as historical data or if examined using different methods.
We do not know how many classic findings survive under contemporary conditions.
So, what action may be needed to keep the approach anchored in reality? First, as we exemplify, the assumptions under which software-engineering events are generated are changing. We urgently need
to create methods that work under new assumptions (e.g., redefining measurement constructs to account for AI-assisted commits, or developing new metrics that capture human effort in the presence of AI assistance).
This includes detecting the types of assumption violations and irregularities listed in Table~\ref{tab:assumptions}, as well as new ones that yet remain to be identified.
Second, we need a broad and systematic replication agenda to re-analyze established constructs using contemporary methods on past and present projects, focused specifically on constructs that measure human behavior.
In Table~\ref{tab:replications}, we outline examples of specific replications. Each example targets a finding about human behavior and asks whether it survives when the human-origin assumption is relaxed. Such replications enable \emph{measurement validity research}: determining which constructs still have valid operationalizations and which need entirely new ones.
A systematic replication agenda would catalog influential findings, identify ones that could benefit from
newer methods or might be affected by new data-generation mechanisms, and subject them to \mbox{replication with contemporary methods and contemporary data.}

The traditional obstacle to replications is cost, as a thorough replication might occupy a graduate student for months.
LLMs can alter this calculus. AI assistants such as ChatGPT or Claude can read papers, extract methodological details, and implement analysis pipelines with speed that humans cannot match~\citep{xu2026scaling,kohler2026read}.
An AI-assisted workflow could: (1)~ingest papers and extract research questions, methods, and assumptions (e.g., about commit \mbox{granularity});
(2)~implement and adapt methodologies for modern data; (3)~execute replications at scale across thousands of repositories; (4)~synthesize results identifying where findings replicate or fail.
Human researchers remain essential to validate implementations, interpret discrepancies, and assess whether adaptations preserve the spirit of original analyses.
But AI assistants enable systematic evaluation at a scale that was previously infeasible, as we were limited by human resources.
Beyond replication, AI can assist prospectively by identifying assumptions that may be violated by modern practices and suggesting sensitivity analyses.
To that end, we need to define units of replication, which findings to re-analyze (e.g., code authorship attribution, productivity, collaboration patterns), and what controls
to put in place to ensure that AI replicates the original analysis as faithfully as possible,
while adjusting for changes in data generation and methodological advances.

Limitations must be acknowledged: AI may misinterpret methodological details, cannot access proprietary data, and may introduce biases. Therefore, human assessment remains essential~\citep{Terragni2025}. Nevertheless, the combination of accumulated findings, methodological advances, and AI tools creates conditions for comprehensive reassessment of what software-engineering research knows.

Beyond a replication agenda, the research community should advocate that development tools and platform owners (e.g., GitHub, GitLab) generate \emph{provenance metadata} as a first-class part of the commit record: Was this commit human-authored, AI-generated, or AI-assisted? Which AI model was used? What was the prompt? What was the human's review decision? Such metadata would let the field continue its tradition of adapting to new trace-generating technologies, as it has done for six decades,
and would enable transparency, traceability, and accountability in software devel\-opment, fostering trust in the software supply chain and enabling researchers to understand how AI is shaping development practices.

\section{Conclusion}
Data and methods of empirical software engineering have both transformed over decades---yet we have not systematically examined whether findings established under earlier conditions
still hold. Automation using the same tools as human developers
makes the
assumption that we analyze events of a human actor rapidly obsolete. This reflection is a call for action: The community should prioritize the development of techniques
to replicate classic software measurement studies using contemporary methods on contemporary data, leverage AI-assisted tools to make such replication feasible at scale, and treat the results---whether confirmatory or contradictory---as valued contributions that strengthen our collective understanding.


\begin{acks}
We received funding from the National Institutes of Health (NIH) under project no.\ 1R01GM164731-01, from the Alfred P.\ Sloan Foundation under grant no.\ g-2025-25171, and from Schmidt
Sciences.
\end{acks}

\balance
\section*{Data Availability Statement}
The scripts used to collect the descriptive statistics on squash-merging prevalence presented in Section~\ref{sec:squash-merge}, as well as more details and plots showing the adoption of squash-merging over time, are available on Zenodo~\cite{bock_2026_21727085}: \; \url{https://doi.org/10.5281/zenodo.21727085}

\balance
\bibliographystyle{ACM-Reference-Format}
\bibliography{literature.bib}
\end{document}